# Twinning induced by elastic anisotropy in FCC crystals

Jie Huang[a], Mingyu Lei[a], Guangpeng Sun[a], Guochun Yang [a,b,*], Bin Wen [a,*]

[a]State Key Laboratory of Metastable Materials Science and Technology, Yanshan University, Qinhuangdao 066004, China.

[b]Key Laboratory for Microstructural Material Physics of Hebei Province, School of Science, Yanshan University, Qinhuangdao 066004, China.

*Corresponding authors.

E-mail addresses: yanggc468@nenu.edu.cn (Guochun Yang), wenbin@ysu.edu.cn (Bin Wen).

**Abstract** Dislocation slip and deformation twin are widely regarded as two important mechanisms of active competition in the process of plastic deformation. Calculating and comparing the critical resolved shear stress (CRSS) of two deformation modes are the key to discussing the mechanical properties reflected by different mechanisms in crystals. Here, the paper proposes a model to predict the CRSS of discrete twins, resembling thin layers, using the elastic anisotropy theory and a macroscopic energy perspective. In addition, the directionality of deformation twinning is also verified. We investigated twinning in FCC crystals to illustrate the methodology, and predicted the CRSS of twinning under different variables such as temperature and strain rate, both of which were in excellent agreement with experimental and other theory results. It draws the conclusion that we can promote twinning nucleation by applying shear stress along the <112> direction to reduce the interface energy as a resistance term and increase the difference in strain energy for twinning nucleation. This conclusion provides a guiding direction for exploring and accurately predicting the conditions of twinning in FCC crystals in future.

**Key words:** Twinning, elastic anisotropy, dislocation slip, FCC crystals, CRSS, temperature,

strain rate

**1. Introduction**

    Twinning and dislocation slip are two important complementary mechanisms governing the plastic deformation of materials. The activation of these mechanisms determines the mechanical behavior of crystalline materials. The appearance of twins can effectively improve the mechanical properties of materials, making them highly applicable in various industry. Up to date, massive efforts have been dedicated to further improve their strength by twinning. Deformation twins have been well-known in body-centered cubic (BCC) [1], hexagonal close-packed (HCP) [2-5] and other crystals with lower symmetry [6]. The occurrence of twinning in face-centered cubic (FCC) metals and alloys has become more prevalent since Blewitt et al. [7] reported twins in Cu single crystals during a tensile test at very low temperatures. This phenomenon is now frequently observed in many FCC metals and alloys. The abnormal situation of twinning under the premise of more slip systems has captured the attention of researchers [8-11], so the exploration of FCC twinning has become an increasingly intriguing research direction.

    There exists a direct competition between dislocation slip and deformation twin in FCC crystals because they are generated by operating on the same set of dislocation slip mechanism. The prevalence of one mechanism over the other is closely related to the landscape of generalized stacking fault energy (GSFE) as described by Vitek [12]. According to generalized planar fault energy (GPFE), numerous investigations [13-15] have been conducted to gain a deeper understanding of the twinning mechanism in FCC crystals. Tadmor and his colleagues [16] gave a pioneering description of the competition between

dislocation slip and deformation twin. Basing on their research many researchers [17-21] have put forward various twin parameter criteria. While these criteria provide a basic explanation for the competition between the deformation mechanisms in FCC crystals, they also exist certain limitations. They predominantly focus on the twinning characteristics of the material itself, but pay little attention to the details of the twinning formation process. The critical resolved shear stress (CRSS) can be regarded as a crucial parameter to determining the deformation mode of materials. Recently, a dislocation-based temperature-dependent hardness model is proposed by Feng et al. [22]. Furthermore, many researchers [14, 23, 24] have introduced dislocation-based nucleation models to predict the CRSS of twinning. However, the lack of models for the complete formation of twin crystals under different experimental conditions, especially in extreme environments.

In this work, in order to better understand deformation twin in FCC crystals, we studied the change process of macro energy and proposed a new model of CRSS for twin nucleation. In this model, no dislocation mechanism is involved for explaining the formation of twin boundary. With this model, we can efficiently predict and analyze different influencing factors (such as temperature and strain rate) on the CRSS of twinning in FCC crystals. By applying the model, we conducted a straightforward assessment of the impact of different factors on both dislocation slip and deformation twin mechanisms. Our analysis utilized copper, gold, silver, nickel and aluminum as examples. The comparative analysis sheds light on the distinct effects of these factors on the two deformation mechanisms. In scenarios where the deformation twin mechanism takes precedence over dislocation slip in materials, not only does this streamline the preparation process for nanotwin materials, but it also serves as a valuable reference for exploring FCC crystals with enhanced properties.

## 2. Computational details

### 2.1 First principles calculation method

Vienna ab initio simulation package (VASP) within the density functional theory (DFT) [25] framework was employed to investigate stable and unstable defect energies in our calculations. The specifics of the DFT calculation (method, convergence studies, supercells, etc.) are detailed below. The projector augmented wave (PAW) in the form of Perdew-Burke-Ernzerhof (PBE) [26] was adopted to describe the exchange-correlation functional. The single-particle Kohn-Sham wave function was expanded using the plane waves with a cut-off energy set at 400 eV. The irreducible edge of Brillouzin zone was sampled using a regular Monkhorst-Pack scheme [27] for the calculation of mechanical properties. Convergence criteria for optimization on atomic internal freedoms were selected as differences in the total energy within $1\times10^{-4}$ eV/atom, and the Hellmann-Feynman force within 0.01 eV/Å. We used $11\times11\times1$ and $11\times11\times9$ k-point meshes to calculate the CTB energy and strain energy. A vacuum layer of 10 Å was embedded into supercell along the direction perpendicular to the twin plane to avoid the interaction between the slab and its periodic images in the calculation of CTBEs. Consistent with the above method, we determined the system energy and CTBEs under different strains. To achieve this, we strained the crystal through incremental simple shears, calculating the energy (E) and relaxed volume (V) as a function of strain. The adjustment coefficient is $1+\varepsilon$, where $\varepsilon$ represents the magnitude of shear strain. We also use the DFT code VASP when calculating the energy of different crystal structures. The pseudopotential was derived from the generalized gradient approximation (GGA) method, with the exchange related energy functional of PBE adopted. At this time, the redefined crystal structure is used.

## 2.2 Calculation models and theory

While the directivity of twins is known from the perspective of crystal geometry [28], we also strictly demonstrate twinning directionality from energy. Fig. 1 provides a schematic representation of this perspective. Fig. 1(a) illustrates the process of direction change in the crystal when twinning occurs. The model assumes that the FCC crystal (width, length and height L) has {111} twin plane, as depicted in Fig. 1(b). When shear stress is applied, the crystal will undergo shear deformation. The crystal volume was kept constant during shear deformation without taking the inner displacement into account. This force leads to the formation of twin nucleus, assumed to be cuboids with length x, width x and height h. We consider the formation of twin nucleus in crystals from the perspective of crystal system energy. The energy change in this process is mainly reflected in two main aspects: driving force and resistance. As the crystal deforms under the action of shear stress, the potential energy will also change. When applied slowly, the kinetic energy of the system can be ignored. Just as the consumption of other energy is omitted, the change of potential energy will be transformed into the difference in strain energy between the begin and end states, then cause the crystal to deform. Considering the nonlinear stress-strain relationship of materials in same direction, we calculate the energy of system under different strain variables. The shear strain $\varepsilon$ in the same direction is constrained, assumed to be uniform throughout the crystal. Therefore, the driving force of the whole twin formation is the difference of strain energy between the two opposite directions, which coincides with the mentioned elastic anisotropy along the $\eta_1$ direction in materials. The resistance energy of twin formation is most sensitively influenced by the twin boundary energy. Coherent twin boundaries (CTB) are closely related to stacking faults or the interfacial energy in different crystal structure. The boundary between the twin region and the parent phase region does not necessarily coincide with the twin plane, leading to incoherent twin boundary (ITB). ITBs can be regarded as the boundary wall composed of a series of dislocations [29-31]. Its energy is higher than that of CTB. The energy change formula in this process is as follows:

$$W = W_S - W_V = 2\gamma_{CTB}x^2 + 4\gamma_{ITB}xh - \big(\Delta E_S(\tau) - \Delta E_H(\tau)\big)hx^2, \tag{1}$$

By considering the equal energy of each surface, we can establish the relationship between twin nucleus sizes. We use energy to differentiate the twin size x, then obtain the critical size of twin nucleus and the energy barriers that need to be overcome for nucleation, which are $x_C = \frac{4\gamma_{ITB}}{\Delta E_S(\tau) - \Delta E_H(\tau)}$ and $\Delta G = \frac{32\gamma_{CTB}\gamma_{ITB}^2}{\big(\Delta E_S(\tau) - \Delta E_H(\tau)\big)^2}$, respectively.

In the Fig. 1(c) schematically shows the stacking fault energy curve of FCC crystal along the <112> direction in {111} plane. Notably, even when atoms move in two opposite directions, the final crystal structures remain the same, indicating that the intrinsic stacking fault energy $\gamma_{ISF}$ in the $[11\bar{2}]$ and $[\bar{1}\bar{1}2]$ directions is consistent. However, the unstable stacking fault energy $\gamma_{USFE}$ in the $[\bar{1}\bar{1}2]$ direction is significantly higher than that in the $[11\bar{2}]$ direction [32]. This difference arises from the unfavorable stacking encountered when the crystal moves relatively along the $[\bar{1}\bar{1}2]$ direction. Changes in crystal structure and stacking order at different positions are also marked in the diagram. Based on this, we observe that $[11\bar{2}]$ and $[\bar{1}\bar{1}2]$ directions represent the soft and hard directions, respectively. The Fig. 1(d) illustrates the variation trend of strain energy in hard and soft directions with stress in crystal. We calculate the shear stress by taking the derivative of the energy with respect to the strain [33], as shown in supplementary materials Fig. 1. We can find that when stress is equal, the energy of hard direction is higher than that of soft direction from Fig. 1(d). Consequently, when subjected to shear forces along the hard direction, the crystal undergoes internal energy storage. Upon reaching a critical point, $[11\bar{2}]$ direction twin appears in crystals. Fig. 1(e) shows the variation trend of system energy difference with twin length under different external forces. The energy barrier required for twin nucleation is significantly higher than that for twin growth. As external force increases, the size of twin formation decreases and the energy barrier required to overcome decreases accordingly.

According to the Arrhenius formula [34] and the understanding of rate [35], the temperature and strain rate effect factors on deformation can be further considered.

$$\dot{\varepsilon} = \varepsilon \frac{df}{dt} = \frac{S}{M} \frac{\gamma_{CTB}\dot{N}v^3t^3}{\gamma_{ITB}} e^{-\frac{\gamma_{CTB}\dot{N}v^3t^3}{4\gamma_{ITB}}} \quad (2)$$

where f is the volume fraction of twins (as shown in supplementary materials), ε is the strain due to twinning, S is the twin shear, M is the Taylor factor, it is 3.06 for FCC metals [36].

## 3. Results and discussion

After discussing the theory of modeling, how should we determine which of slip and twinning is dominant? Calculating and comparing the CRSSes of two deformation modes are the key to solve this problem. The competition between deformation twins and slip is influenced by the magnitude and direction of applied stress. Twin nucleation typically requires a rather high and localized stress level, while twin growth demands a much lower stress level. In addition, once nucleated, the deformed twins can thicken in a very short time. Therefore, we regard twin nucleation as the judgment condition for the formation of twins.

Nucleation is a necessary process of twin formation, which can be divided into homogeneous nucleation and heterogeneous nucleation. Heterogeneous nucleation is usually formed by defects such as existing grain boundaries or dislocations, generally has a lower energy barrier compared to homogeneous nucleation. Given that FCC crystals commonly have defects, we primarily consider heterogeneous nucleation, focusing on two coherent twin boundaries (CTBs) and one incoherent twin boundary (ITB). In the context of this study, the single shear stress τ in the $[11\bar{2}]$ direction serves as the CRSS for twinning. Feng [22] proposed a formula for calculating the CRSS of dislocation slip in materials. We can solve the CRSS required for twinning nucleation by using the energy change formula mentioned above. It is well known that the most common twin and slip systems are the $\{111\}[11\bar{2}]$ and $\{111\}[1\bar{1}0]$ in FCC materials. The geometric parameters required for calculating CRSS are also shown in Table 1. The activation energy changes with the applied stress for twinning and dislocation slip in five metals are shown in Fig. 2(a). The intersection with x axis is the CRSS of each deformation mode. We can clearly observe that the two deformation modes of the five materials have obvious intersections. Generally, it is believed that the activation energy of deformation twin nucleation is greater than that of slip at low to medium stress levels. The CRSSes required for Cu, Au, Ag, Ni and Al twinning is maintained at about 128, 94, 69, 251 and 217 MPa, respectively. In Fig. 2(b), we also discussed the relationship between the axial ratio of twin nuclei and stress. We found that all values were less than 1, which proves that the formed twin nuclei have a flat structure. In addition, the axial ratio of all materials gradually decreases with increasing stress, and their slopes are consistent with the order of CRSS for twinning. The Fig. 3(a) shows the relationship between twin volume fraction and time in Cu. With the increase of time, the twin volume fraction shows an S-shaped curve gradually increasing trend. When the applied external force is increased, the time required to form twins is significantly shortened. The trend of other materials is consistent with Cu. Then analyze the competition of twinning and dislocation slip by different influence factors, as described later.

**3.1 The effect of temperature**

Twinning is investigated as a dissipative process outside thermodynamic equilibrium [37]. Additionally, the diffusion rate increases at high temperatures, which is not necessary to promote twinning, unlike the conditions for slip under many circumstances. Fig. 3(b) illustrates the deformation mode and corresponding CRSS of five FCC crystals at different temperatures. The CRSSes for both dislocation slip and deformation twin decreases with increasing temperature, albeit with distinct degrees of change. According to the results of the transformation of deformation mode affected by temperature discussed above, we can find that dislocation slip in materials is extremely sensitive to temperature. However, temperature has a comparatively smaller effect on the stress required to induce twinning compared to that needed for slip. These results are in good agreement with the experimental results [38-41] and calculation results of others [14, 42].

The CRSSes for the two deformation modes exhibit an intersection with varying temperature, defining the transition temperature. Above the transition temperature, the CRSS of slip is less than that of twinning, leading to crystal deformation primarily through dislocation slip. Conversely, below the transition temperature, the CRSS for twinning is smaller, resulting in the appearance of twins in the crystal. It is worth noting that once reach the CRSS of twinning, twin nucleation can appear in the material. Therefore, deformation twins are more likely to occur at lower temperatures, a trend consistent with experimental findings [43, 44].

**3.2 Effect of strain rate**

The strain rate emerges as a prominent influencing factor on the choice of deformation mode in FCC crystals. In the realm of crystallography, slip and twinning are the two primary mechanisms by which atoms rearrange to accommodate external deformation. As the strain rate increases, the internal strain within the crystal also intensifies. This can lead to an escalated competition between slip and twinning. Here, based on the intersection point of previous temperature, we attempt to compare the CRSSes of twin and slip in five materials under different strain rates, considering a range of $10^{-3}$-$10^{3}$ s$^{-1}$. When considering strain rates, we assume that the temperatures of Cu, Au, Ag, Ni and Al are 100 K, 80 K, 100 K, 100 K and 30 K, respectively. It can be seen from the Fig. 3(c) that the deformation mode of five materials all changes from slip dominant to twin dominant with increasing strain rate. In short, twinning is easier in the FCC crystals at high strain rate. In twinning, atomic planes undergo relatively large displacements compared to the rest of the crystal, making it advantageous under conditions of faster strain rates.

In addition, the competitive effects of temperature and strain rate on slip and twinning deformation modes in FCC crystals are complementary. There is a strain rate with the change of deformation mode in the competition between slip and twinning. The position of the transition point is also related to temperature, as shown in Fig. 3(d). It illustrates the impact of strain rate on the transition temperature to melting point ratio for slip and twinning deformation modes in five different materials using a bar chart. It is evident that, regardless of the material, there is a general trend of the transition temperature gradually increasing with the higher strain rate. When comparing the transition temperatures of the five materials at a consistent strain rate, the sensitivity to temperature follows the order: copper, silver, nickel, gold, and aluminum. This result aligns with the expectation that aluminum is least prone to twinning among the materials studied. The reasons for this order are primarily twofold. First, the significant difference in their stacking fault energies, coupled with distinct trends with temperature variation, plays a crucial role. However, in this model, the resistance to twin nucleation primarily stems from interfacial energy, leading to variations in the competitive process. Second, the sensitivity of their shear modulus to temperature varies, and in formula we employed to calculate the CRSS for slip, this parameter holds substantial importance. In essence, it is the differential impact of temperature on the required parameters for each model that gives rise to this order.

## 4. Conclusions

When discussing how the dislocation slip and deformation twin modes of FCC crystals compete, the CRSS serves as a very important criterion. In summary, we solved the formula of CRSS for twinning according to the macroscopic energy change of twinning process in FCC crystals. This formula was further extended by incorporating the Avrami equation and Arrhenius formula. We extensively discussed the impact of temperature and strain rate on the mechanical properties of FCC crystals. We find that factors both have two modes of mutual exchange conversion points, leading to the conclusion that twins are more likely to appear at low temperature and high strain rate in FCC crystals. In Fig. 4, the dominant deformation mode is marked in the graph. Our results not only provide a new way to analyze the competitive relationship of deformation modes in FCC crystals, but also help to better guide the emergence of twins in FCC crystals in experiment.

## 5. Acknowledgments

This work was supported by the National Natural Science Foundation of China (Grant Nos. 51925105, 51771165) and the National Key R&D Program of China (YS2018YFA070119).

**Table 1** Geometric parameters required for calculating CRSS

| Materials | ITB | $n$ | $v_a$ (THz) | $G$ (eV) |
|-----------|-----|------|-------------|----------|
| Cu | 498 | 8.47 | 7.21 | 0.74 |
| Al | 357 | 6.03 | 9.65 | 0.65 |
| Ni | 588 | 9.14 | 6.27 | 1.02 |
| Au | 94.5 | 5.89 | 2.75 | 0.79 |
| Ag | 397 | 5.86 | 4.93 | 0.64 |

The units of ITB and n are mJ·m$^{-2}$ and $10^{28}$ atom/m$^3$, respectively. The unit of $v_a$ and $G$ are THz and eV, respectively.

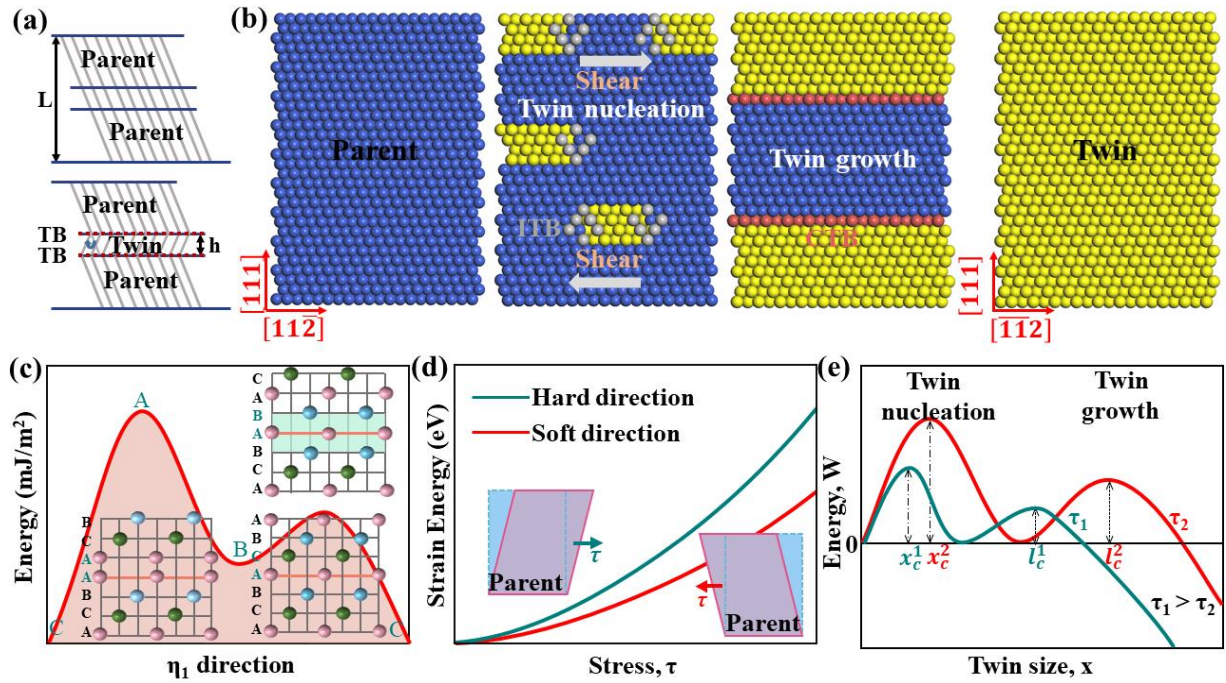

**Fig. 1** The anisotropy of FCC crystals. (a) illustrates the process of direction change in the crystal when twinning occurs. The arrow indicates change in crystal orientation from parent to twin. (b) represents the complete process of twinning nucleation and growth when shear stress is applied. (c) is stacking fault energy curve along $\eta_1$ direction. It schematically shows two opposite FCC twinning modes giving identical final twin configurations of $(111)[11\bar{2}]$ and $(111)[\bar{1}\bar{1}2]$ directions. We also illustrate the crystal structure and stacking order of atoms moving in different directions. The green shadow represents the stacking fault in FCC crystals. (d) is the variation trend of strain energy of soft and hard directions with stress in FCC crystals. (e) is the relationship between the change of system energy and the length of twinning during the formation process of twinning under different external forces. We also marked the energy barrier changes during twin nucleation and growth.

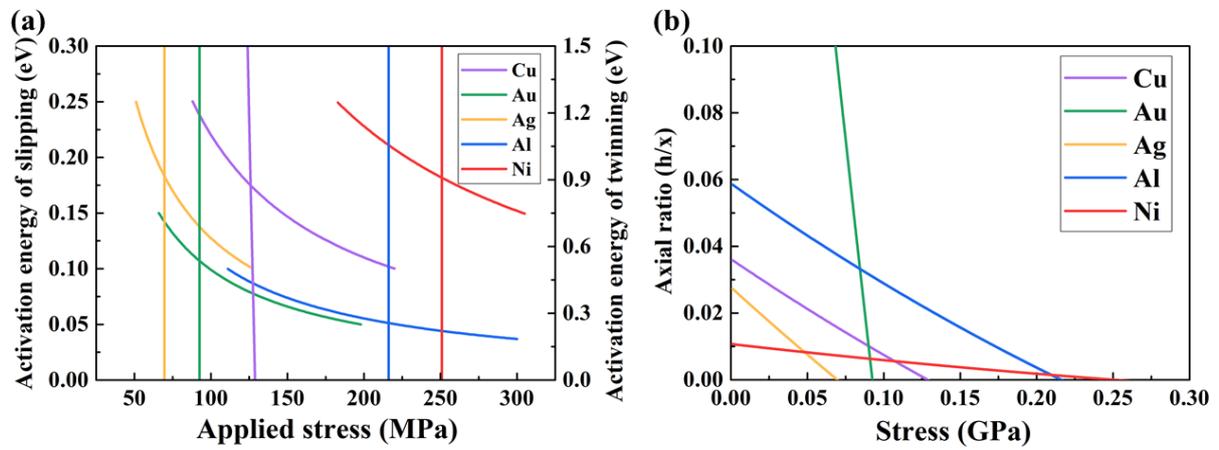

**Fig. 2** Parameters in twinning nucleation process. (a) is the curve of the activation energy of two deformation modes as a function of external force. (b) represents the relationship between twin nucleation axial ratio and stress.

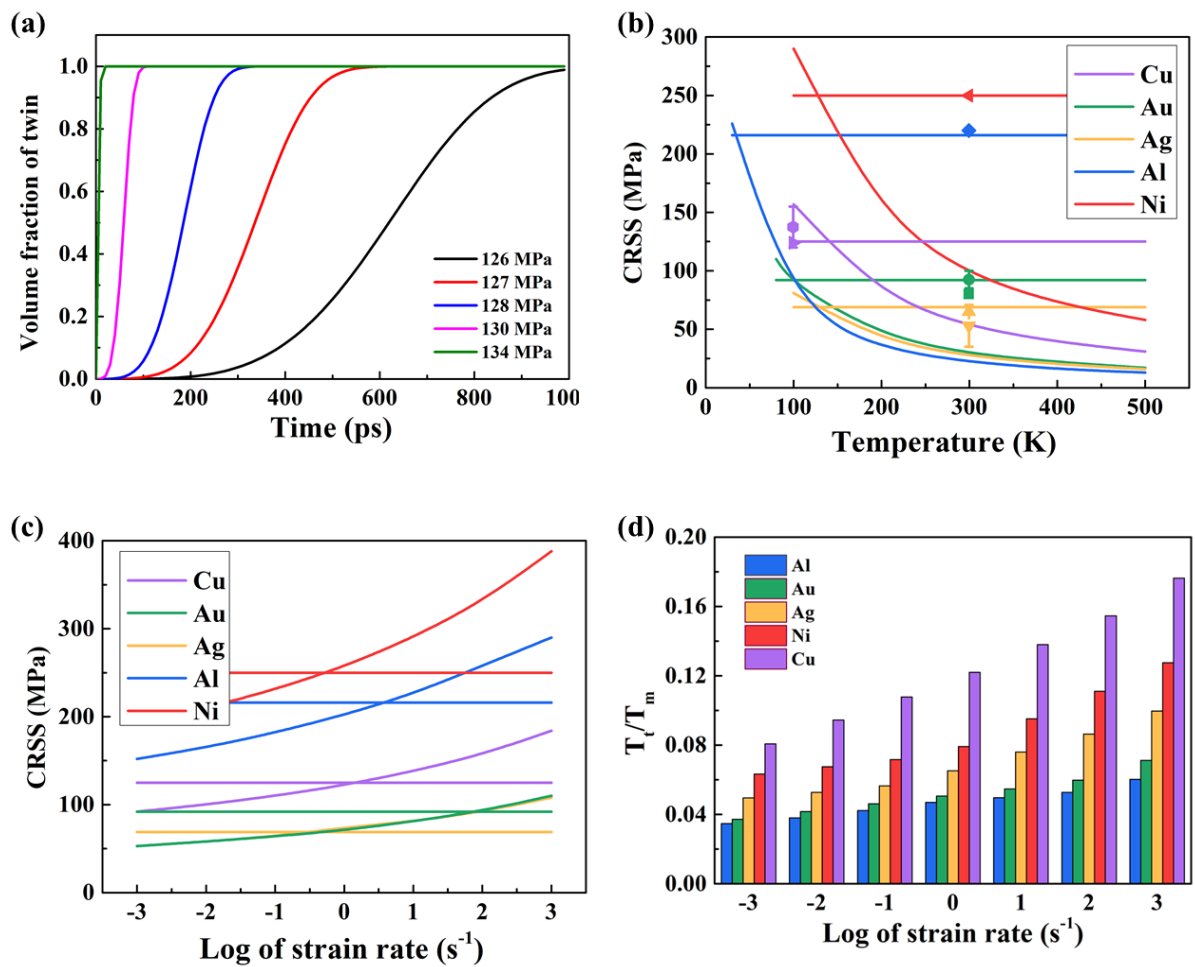

**Fig. 3** The competition between slip and twinning. (a) represents the relationship between volume fraction of twin and time under different stress conditions in Cu. (b) shows the effect of temperature on the CRSS for slip and twinning in five materials. (c) represents the effect of strain rate on the CRSS for slip and twinning. (d) is the influence of strain rate on the transition temperature to melting point ratio of slip and twinning deformation modes.

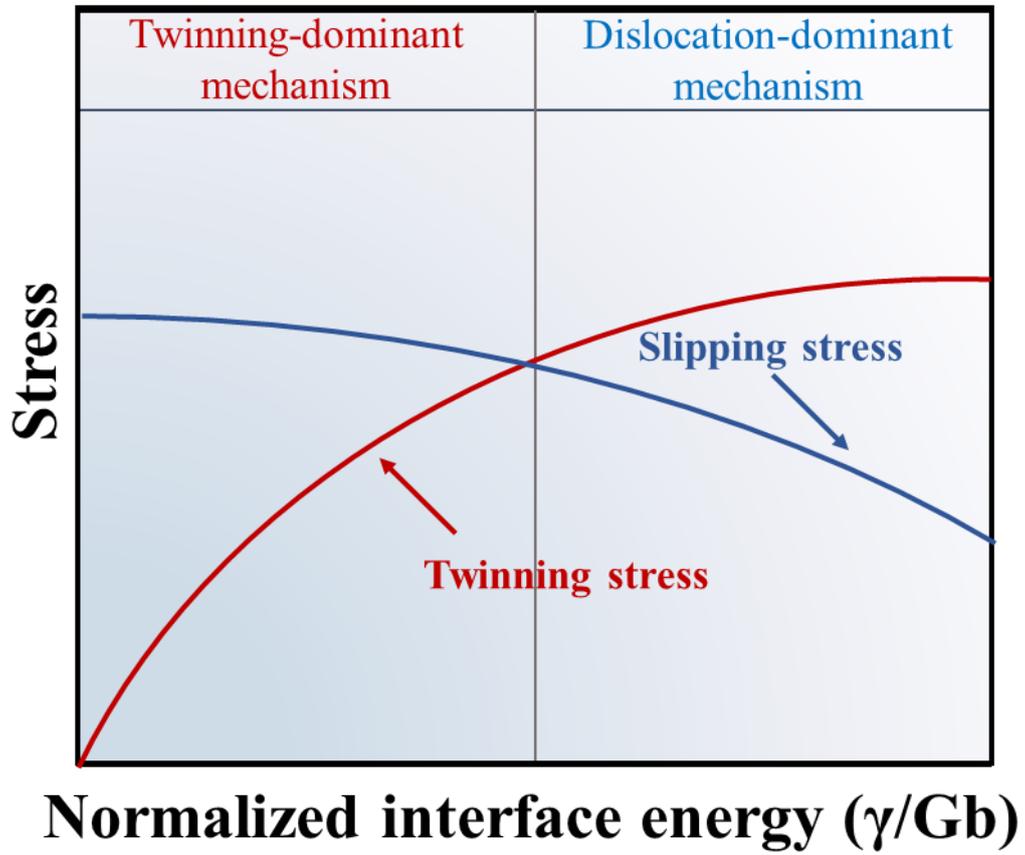

**Fig. 4** Schematic illustration of two different mechanisms.